\documentclass[pdflatex,Numbered,sn-basic]{sn-jnl}

\usepackage{hyperref}
\usepackage{graphicx}
\usepackage{amsmath}
\usepackage{macros}
\usepackage{stmaryrd}
\usepackage{amssymb}
\usepackage{dirtytalk}
\usepackage[inline]{enumitem}
\usepackage{multicol}
\usepackage{booktabs}
\usepackage{multirow}
\usepackage{graphicx}
\usepackage[linesnumbered,noline,noend]{algorithm2e}
\usepackage[utf8]{inputenc}
\usepackage{booktabs}
\usepackage{array}
\usepackage{subcaption}
\usepackage{hyperref} 

\theoremstyle{thmstyleone}%
\newtheorem{theorem}{Theorem}
\theoremstyle{thmstyletwo}%
\newtheorem{lemma}{Lemma}%

\theoremstyle{thmstylethree}%

\begin{document}

\title[Solving Fuzzy Satisfiability via Mixed-Integer Non-Linear Programming]{Solving Fuzzy Satisfiability via Mixed-Integer Non-Linear Programming}


\author*[1,2]{\fnm{Pablo F.} \sur{Castro}}\email{pcastro@dc.exa.unrc.edu.ar}

\affil*[1]{\orgdiv{Department of Computing}, \orgname{Universidad Nacional de Río Cuarto}, \orgaddress{\street{Ruta Nac. 36 Km. 601}, \city{Rio Cuarto}, \postcode{5800}, \state{Córdoba}, \country{Argentina}}}

\affil[2]{\orgname{CONICET}, \orgaddress{\street{Godoy Cruz 2290}, \city{Buenos Aires}, \postcode{C1425FQB}, \state{CABA}, \country{Argentina}}}


\abstract{This paper introduces an algorithm for SAT solving for fuzzy logics. In contrast to the Boolean case, for which numerous SAT solvers exist, the SAT problem for fuzzy logics has attracted less attention, even though these tools have interesting applications.
Unlike existing SAT solvers for fuzzy logics, our algorithm uses  MINLP (Mixed Integer Non-Linear Programming) solvers to check the satisfiability of fuzzy formulas. This approach offers certain benefits; for instance, our tool can handle all major variations of fuzzy propositional logic, whereas other fuzzy solvers are usually tailored to specific versions of fuzzy logic. We implemented the algorithm in a tool (\textsf{SATFuL}) and conducted experiments, showing that the performance of our tool is comparable to state-of-the-art fuzzy solvers for {\luk} logic and outperforms available solvers for product logic. The approach is sound and complete and can be easily extended to accommodate new fuzzy operators.}


\maketitle

\section{Introduction}
\emph{Boolean SAT solvers} \cite{Knuth2015}---tools used to verify whether a Boolean formula is satisfiable---have important applications in various areas of computer science. Some notable examples include \emph{SMT solvers} \cite{Barrett2009SMT}, which rely on SAT solvers for efficient Boolean reasoning, and \emph{bounded model checkers} \cite{DBLP:journals/ac/BiereCCSZ03}, which reduce the problem of verifying a temporal formula to  Boolean satisfiability problems.

The satisfiability problem is also of interest in \emph{fuzzy logics} \cite{Hajek1998}---logics in which formulas are evaluated in the interval $[0,1]$--- sometimes referred to as \emph{infinite-valued logics}. These logics have interesting applications in various areas of computer science, including reasoning about neural networks \cite{DBLP:series/sist/NolaLV16}, image processing \cite{Hassan09}, or  multi-agent systems \cite{Marchioni2019}. However, the development of SAT solvers for these kinds of logics has received relatively  less attention from the computer science research community. One reason for this is that satisfiability methods in fuzzy logic are highly dependent on the arithmetical interpretation of logical operators. In fact, there are several variants of fuzzy logics, some of the most prominent are: \emph{G\"odel logic}, \emph{product logic}, and \emph{{\luk} logic}. In Section~\ref{sec:prel}, we give a brief introduction to these formalisms.

Some approaches to solving the SAT problem for fuzzy logics are worth mentioning.
In \cite{DBLP:journals/jar/SchockaertJV12}, the authors propose to reduce the problem of satisfiability of infinite logics to that of finite-valued logics, allowing constraint solvers to be used for checking satisfiability. This approach can handle formulas of reasonable size but suffers from scalability issues for certain formulas, e.g., when they are unsatisfiable. In \cite{Brys2012}, a different approach is proposed, using \emph{evolution strategies} to solve the satisfiability problem in fuzzy logics. This method is incomplete and exhibits performance comparable to the previous one, with some improvements for specific classes of formulas.
For product logics, almost no SAT solvers are available. An exception is \textsf{MNiBLoS} \cite{Vidal2016}, which leverages the SMT solver \textsf{Z3} \cite{DeMoura08} to check satisfiability. However, this approach is based on an incomplete transformation of nonlinear arithmetic problems into negative arithmetic. As a result, \textsf{MNiBLoS} may incorrectly classify unsatisfiable clauses as satisfiable.

Furthermore, the performance of existing fuzzy SAT solvers remains far behind that of Boolean SAT solvers, which can typically handle formulas several orders of magnitude larger. In addition, Boolean SAT solvers offer advanced features such as incremental solving, unsatisfiability core extraction, and other capabilities that are generally absent from the aforementioned fuzzy solvers.

In this paper, we build on the initial ideas proposed in \cite{Hahnle2003} to reduce the problem to MILP (Mixed-Integer Linear Programming) problems. This approach was implemented by a few tools in the area; the primary reason appears to be the scalability issues of MILP solvers at that time. However, in recent years, MILP solvers have shown remarkable advances. Commercial MILP solvers such as \textsf{Gurobi} \cite{gurobi} can solve linear systems with thousands of variables, and academic solvers like \textsf{SCIP} \cite{BolusaniEtal2024OO} also offer strong performance. Additionally, these solvers have recently been extended to handle nonlinear systems. 
We implement this algorithm in a prototype tool, \textsf{SATFuL}, a SAT solver for fuzzy logics that uses 
MINLP (Mixed-Integer Non-Linear Programming) solvers to check formula satisfiability. The procedure is sound and complete (modulo the correctness of the MINLP solver used). The tool can handle all versions of fuzzy logic mentioned above. The algorithm used by \textsf{SATFuL} is versatile in the sense that it can easily be modified to cope with different logics. We compare our solver with \textsf{fuzzyDL}~\cite{Bobillo2016}, a state-of-the-art solver for description fuzzy logic that supports {\luk} and G\"odel logic, and with \textsf{MNiBLoS}~\cite{Vidal2016}, one of the few solvers for product logic. Furthermore, we implemented the SAT-solving algorithm using the SMT solver \textsf{Yices}~\cite{Dutertre:cav2014}, which supports non-linear arithmetic and conditional expressions, allowing a more direct encoding of the fuzzy operators, and compared its performance with that obtained with numerical solvers.

The structure of the paper is as follows. 
In Section~\ref{sec:prel}, we introduce the basic notions needed for the rest of the paper. Section~\ref{sec:sat} introduces the satisfiability algorithm and discusses its correctness.
Numerical and logical optimization for the algorithm are proposed in Section~\ref{sec:opt}.
We discuss related work in Section~\ref{sec:related}.
In Section~\ref{sec:experiments}, we show an experimental evaluation of the tool. Finally, we draw some conclusions.

\section{Preliminaries} \label{sec:prel}
    In this section, we introduce the concepts and notation required for the remainder of the paper. 
We focus on fuzzy logics, i.e.,  many-valued logics in which the truth value is taken from the interval $[0,1]$. Most fuzzy logics are constructed upon the concept of \emph{t-norm}. A t-norm is a binary operator $\otimes : [0,1]^2 \rightarrow [0,1]$ that satisfies \emph{commutativity}, \emph{associativity}, \emph{monotonicity}, and \emph{identity}. 
There are different choices for $\otimes$; each of them gives a different logic. In this paper, we concentrate on the following logics.
\paragraph{{\luk} Logic} This logic is obtained by considering  $x \otimes y = \max(0, x+y-1)$ as the 
t-norm. Hence, the usual operators are defined as follows:
\begin{multicols}{2}
\begin{itemize}
    \item $x \lukand y = \max(0, x+y-1)$,
    \item $x \lukor y = \min(1, x+y)$,
    \item $x \lukimplies y = (x \leq y) {?} 1 {:} (1-x)+y$
    \item $\luknot x = 1-x$. 
\end{itemize}
\end{multicols}

\paragraph{Product Logic} Product logic is obtained  by using the product as  t-norm, in this case, we obtain the following operators:
\begin{multicols}{2}
\begin{itemize}
    \item $x \productand y = x * y$,
    \item $x \productor y = x+y - x*y$,
    \item $x \productimplies y = \cond{x \leq y}{1}{y/x}$,
    \item $\productnot x = \cond{x = 0}{1}{0}$.
\end{itemize}
\end{multicols}

\paragraph{G\"odel Logic} This logic is obtained by using the t-norm $\min$ as the conjunction and defining the logical operators as follows:
\begin{multicols}{2}
\begin{itemize}
    \item $x \godeland y = \min(x,y)$,
    \item $x \godelor y = \max(x,y)$,
    \item $x \godelimplies y = \cond{x \leq y}{1}{y}$,
    \item $\godelnot x = \cond{x = 0}{1}{0}$.
\end{itemize}
\end{multicols}
Let $\mathcal{X}$ be a (finite) set of fuzzy variables, an inductive definition of fuzzy formulas is direct: a formula is either a fuzzy variable, a (rational) constant in $[0,1]$\footnote{The standard definition of fuzzy logic does not consider constants, this is sometimes called Rational Pavelka logic.}, or the application of any operator of the corresponding logic to formulas.   

Note that many operators can be defined using a set of basic ones. For instance, in {\luk}'s logic, any operator can be defined using $\lukimplies$. Similarly, all the disjunctions can be defined using the corresponding conjunctions and {\luk}'s negation. We refer the interested reader to \cite{Hajek1998} for an in-depth introduction to fuzzy logics.

\paragraph{SAT in   Fuzzy Logics} Given a vocabulary $\mathcal{X}$, a \emph{valuation} is a function $v:\mathcal{X} \rightarrow [0,1]$. Valuations can be recursively extended to formulas using the definitions given above.
We use the notation $[0,1]^{\mathcal{X}}$ to denote the space of all valuations.  
We can restate the SAT problem for many-value logics as follows. We say that a fuzzy formula $\phi$ is $1$-SAT iff there is a valuation $v$ such that $v(\phi)=1$, and we say that $\phi$ is $k$-SAT iff $v(\phi) \geq k$. $1$-SAT and $k$-SAT for the three logics above are  NP-complete \cite{Hajek1998}. This can be generalized to \emph{clauses}, that is, formulas of style $\ell \leq \phi \leq u$ (with $\ell, u \in [0,1]$), we say that this clause is satisfiable, if there is a valuation $v$ such that $\ell \leq v(\phi) \leq u$, denoted $v \vDash_\infty \ell \leq \phi \leq u$. Furthermore, we say that a set of clauses is \emph{satisfiable} if there is a valuation that satisfies all the clauses in the set. In this case, we say that the set of clauses is SAT$_\infty$. We use the notation $\mathit{Var}(\Phi)$ to denote the set of variables occurring in  $\Phi$.

\paragraph{Mixed Integer Linear Programming} Given a collection $\mathcal{X}=\{x_0,\dots,x_{n-1}\}$ of variables over $\mathbb{R}$,
a Mixed Integer Linear Programming (MILP) problem can be described as a tuple $\langle \mathcal{I}, \mathcal{Z}, \sum_{x_i \in \mathcal{X}} x_i * c_i \rangle$, where 
$\mathcal{I}$ is a finite collection of linear inequalities; $\sum_{x_i \in \mathcal{X}} x_i * c_i$ is the objective function to minimize (or maximize); and $\mathcal{Z} \subseteq \mathcal{X}$ is a set of variables that takes integer values. The constraints can be expressed in matrix form as:  $Ax \leq b$, where $A$ is an $n \times m$ matrix of real numbers, $x$ is the (row) vector $\langle x_0,\dots,x_{n-1}\rangle$, and $b$ is a row vector of size $m$ of constants. If $\mathcal{Z}=\emptyset$, then  it is called a \emph{linear programming problem}. 
Mixed Integer Non-Linear Programming (MINLP) problems extend MILP problems with the possibility of having non-linear constraints.
Linear programming is in P, MILP is NP-complete, while MINLP is NP-hard and decidable for bounded problems. Given a problem $P=\langle \mathcal{I}, \mathcal{Z}, \sum_{x_i \in \mathcal{X}} x_i * c_i \rangle$,  a \emph{feasible solution} is defined to be assignment of values to the variables (a function $f \in \mathbb{R}^\mathcal{X}$) such that satisfies  the inequalities $\mathcal{I}$, an \emph{optimal solution} is a feasible solution that minimizes (or maximizes) $\sum_{x_i \in \mathcal{X}} x_i * c_i$.
$\mathcal{F}(P)$ denotes the set of feasible solutions of $P$, while $\mathcal{O}(P)$ denotes the set of optimal solutions of $P$. Furthermore, when there is no objective function to optimize, we just write:
$P=\langle \mathcal{I}, \mathcal{Z}, \emptyset \rangle$; in such a case, any feasible solution is an optimal solution.

Given an assignment $f \in \mathbb{R}^\mathcal{X}$, we denote by $f|_{\mathcal{X}'}$ the restriction of $f$ to a set $\mathcal{X}' \subseteq \mathcal{X}$.
\RestyleAlgo{ruled}
\begin{algorithm}[!h]
\scriptsize 
\SetKwInOut{Input}{Input}
\SetKwInOut{Output}{Output}
\SetKwProg{Fn}{Function}{}{end}

\Input{A fuzzy formula $\phi$}
\Output{Tuple $\langle\mathcal{I}, \mathcal{Z}\rangle$ where $\mathcal{I}$ is a set of inequalities and $\mathcal{Z}$ is a set of integer variables.}

\caption{The \textsc{toMINLP} procedure.}\label{alg:to-minlp}

\Fn{\textsc{toMINLP}($\phi$)}{
    Let $x_\phi \in [0,1]$ be a fresh variable\;
    
    \lIf{$\phi = x$ \textnormal{for} $x \in \mathcal{X}$}{
        \KwRet{$\langle \{ 0 \leq x \leq 1\}, \emptyset \rangle$}
    }
    \lIf{$\phi = k'$}{
        \KwRet{$\langle \{x_{\phi} = k'\}, \emptyset \rangle$}
    }
    
    \If{$\phi = \phi' \# \psi'$ \textnormal{with} $\# \in \{\lukimplies, \productand, \productimplies \}$}{
        $\langle \mathcal{I}_0, \mathcal{Z}_0 \rangle \gets \textsc{toMINLP}(\phi')$\;
        $\langle \mathcal{I}_1, \mathcal{Z}_1 \rangle \gets \textsc{toMINLP}(\psi')$\;
        $\mathcal{Z} \gets \emptyset$\;
        
        \If{$\phi = \phi' \lukimplies \psi'$}{
            Let $x \in \{0,1\}$ be a fresh variable\;
            $\mathcal{I} \gets \left\{
            \begin{aligned}
                &x \geq x_{\phi'} - x_{\psi'}, \\
                & (1-x) \geq x_{\psi'} - x_{\phi'}, \\
                &x_{\phi} = (1-x) + x \cdot (1-x_{\phi'} + x_{\psi'})
            \end{aligned}
            \right\}$\;
            $\mathcal{Z} \gets \{x\}$\;
        }
        \lIf{$\phi = \phi' \productor \psi' $}{
            $\mathcal{I} \gets \{ x_\phi = x_{\phi'}+x_{\psi'} - x_{\phi'}*x_{\psi'}\}$
        }
        \lIf{$\phi = \phi' \productand \psi' $}{
            $\mathcal{I} \gets \{ x_{\phi} =  x_{\phi'} * x_{\psi'} \}$
        }
        \If{$\phi = \phi' \productimplies \psi' $}{
            Let $x,x' \in \{0,1\}$ be fresh variables\;         
            $\mathcal{I} \gets \left\{ 
            \begin{aligned}
                & x \geq x_{\phi'} - x_{\psi'}, \\
                & (1-x) \geq x_{\psi'} - x_{\phi'}, \\
                & x_\phi \geq 1 - x\\
                & x' \cdot (x_{\phi'} - x_{\psi'}) \geq x, \\
                & x_{\psi'} - (1-x') \leq x_\phi \cdot x_{\phi'}, \\
                & x_\phi \cdot x_{\phi'} \leq x_{\psi'} + (1-x')
            \end{aligned} 
            \right\}$\;
            $\mathcal{Z} \gets \{x,x'\}$\;
        }
        \KwRet{$\langle \mathcal{I} \cup \mathcal{I}_0 \cup \mathcal{I}_1, \mathcal{Z} \cup \mathcal{Z}_0 \cup \mathcal{Z}_1 \rangle$}\;
    }
    
    \If{$\phi = \productnot \phi'$}{
        $\langle \mathcal{I}_0, \mathcal{Z}_0  \rangle \gets \textsc{toMINLP}(\phi')$\;
        Let $x \in \{0,1\}, x' \in \mathbb{R}$ be fresh variables\;
        $\mathcal{I} \gets \left\{ 
                              \begin{aligned}
                                 x_{\phi'} \leq 1- x, \\
                                 x'*(x + x_{\phi'}) \geq 1,\\
                                 x_{\phi} = x
                              \end{aligned} \right \}$\;
        
        $\mathcal{Z} \gets \{x\}$\;
        \KwRet{$\langle \mathcal{I} \cup \mathcal{I}_0, \mathcal{Z} \cup \mathcal{Z}_0 \rangle$}\;
    }
}
\end{algorithm}
\section{The SAT Algorithm} \label{sec:sat}
In this section, we present the SAT algorithm used by \textsf{SATFuL}. This algorithm reduces a  $\text{SAT}_{\infty}$ problem to a MINLP problem. The algorithm can deal with any of the fuzzy operators introduced in Section~\ref{sec:prel}. Note that the operators of the G\"odel logic can be expressed using the {\luk} ones, so they are not included in the description of the algorithm, but they can be straightforwardly added. 

\begin{algorithm}[!h]
\small
\SetKwInOut{Input}{Input}
\SetKwInOut{Output}{Output}
\SetKwProg{Fn}{Function}{}{end}

\Input{A set of bounded fuzzy formulas $\Phi$}
\Output{``SAT'' if the underlying MINLP problem is feasible, ``UNSAT'' otherwise}

\caption{The \textsc{SAT} procedure.}\label{alg:mainsat}

\Fn{\textsc{SAT}($\Phi$)}{
    $\mathcal{I} \gets \emptyset$\;
    $\mathcal{Z} \gets \emptyset$\;
    
    \ForEach{$l \leq \phi \leq u \in \Phi$}{
        $\langle \mathcal{I}', \mathcal{Z}' \rangle \gets \textsc{toMINLP}(\phi)$\;
        $\mathcal{I} \gets \mathcal{I} \cup \mathcal{I}' \cup \{x_\phi \geq l\} \cup \{x_\phi \leq u\}$\;
        $\mathcal{Z} \gets \mathcal{Z} \cup \mathcal{Z}'$\;
    }
    
    $P_\Phi \gets \langle \mathcal{I}, \mathcal{Z}, \emptyset \rangle$ \label{alg:minlp-problem} \tcp*{MINLP problem}
    
    \If{\textnormal{feasible}($P_\Phi$)}{
        \KwRet{``SAT''}\;
    }
    \Else{
        \KwRet{``UNSAT''}\;
    }
}
\end{algorithm}

Algorithm~\ref{alg:mainsat} shows the basic procedure. It takes a set of clauses and determines whether the set is satisfiable. It uses the auxiliary procedure Algorithm~\ref{alg:to-minlp}, which translates a clause to a MINLP problem. 
Several aspects of the algorithm are worth noting. First, Algorithm~\ref{alg:to-minlp} takes a formula and produces the main components of a MINLP problem. To do so, the algorithm considers a fresh variable $x_{\phi'}$ for any subformula $\phi'$ appearing in the clauses. This is used to connect the constraints obtained for the components of a clause. Second, Algorithm~\ref{alg:mainsat} uses $\textsc{toMINLP}$ to obtain the constraints corresponding to each clause and adds the inequalities corresponding to the lower and upper bounds. 
%
\subsection{Algorithm Correctness}

To prove the correctness of Algorithm~\ref{alg:mainsat}, first, we introduce some necessary notation.
Given a finite set of clauses $\Phi$, we denote by $P_{\Phi}$ the corresponding MINLP problem (line~\ref{alg:minlp-problem} of Algorithm~\ref{alg:mainsat}). We extend this notation to formulas and, given a formula $\phi$, $P_\phi$ denotes the corresponding MINLP problem obtained by applying Algorithm~\ref{alg:mainsat} to the clause $0 \leq \phi \leq 1$. 
Note that any assignment $f \in \mathbb{R}^\mathcal{X}$ such that $0 \leq f(x) \leq 1$ (for all $x \in \mathcal{X}$) is also a valuation over $\mathcal{X}$. In the following, $\mathit{Var}(\phi)$ denotes the set of fuzzy variables appearing in formula $\phi$, and similar notation is used for clauses.

The following lemma proves that any assignment (say $f$) satisfying the MINLP problems constructed by Algorithm~\ref{alg:mainsat} for formula $\phi$, assigns to the indexed variable $x_{\phi'}$ (being $\phi'$ a subformula) the same value as the valuation $f|_{\mathcal{X}} \in [0,1]^\mathcal{X}$, which is recursively defined as in Section~\ref{sec:prel}.

\begin{lemma}\label{lemma:MINLP-to-SAT} Let $\psi$ be a formula of {\luk}, G\"odel, or product fuzzy logic, and let   $\mathcal{X} = \mathit{Var}(\psi)$. For all subformulas $\phi$ of $\psi$ it holds:
\[
\forall f \in \mathcal{F}(P_\psi) : f(x_{\phi}) = f|_{\mathcal{X}}(\phi)
\]
\begin{proof}
The proof is by induction on $\phi$.

\noindent \textbf{Base Case:} For all the logics we have two base cases: $\phi = k$ (for a constant $k$) or $\phi=x$ (for a fuzzy variable). In the first case we have that
$f(k) = k = f|_\mathcal{X}(k)$ by definition. In the second case we have that $x_\phi = x$ and so $f(x) = f|_\mathcal{X}(x)$.

\noindent \textbf{Inductive Case:} We consider the possible cases for each logic:
Case $\phi = \phi' \productand \psi'$. 

By definition of $P_\psi$ and taking into account that $f \in \mathcal{F}()$ we have $f(x_{\phi' \productand \psi'}) = f(x_{\phi'}) * f(x_{\phi'})$, by induction we have $f(x_{\phi' \productand \psi'}) = f|_{\mathcal{X}}(\phi') * f|_{\mathcal{X}}(\psi')$ and by definition of $\productand$ we obtain:
$f(x_{\phi' \productand \psi'}) = f|_{\mathcal{X}}(\phi' \productand \psi')$.

Case $\phi = \phi' \lukimplies \psi'$: 
Since $f \in \mathcal{F}(P_{\psi})$ we have that the following equations hold: 
\begin{enumerate}
    \item \label{form1} $f(x) \geq f(x_{\phi'}) - f(x_{\psi'})$,
    \item \label{form2} $1-f(x) \geq f(x_{\psi'}) - f(x_{\phi'})$,
    \item \label{form3} $f(x_\phi) = (1-f(x))  + f(x) \cdot(1 - f(x_{\phi'}) + f(x_{\psi'}))$.
\end{enumerate}
We proceed by cases. If $f(x_{\psi'}) > f(x_{\phi'})$, then $f(x) = 0$ because of (\ref{form1}) and taking into account that $x$ is binary, and then we have: 
\begin{equation*}
\begin{aligned}
f(x_\phi) &= 1 \\
&= \begin{cases} 
1 & \text{if } f(x_{\phi'}) \leq f(y_{\psi'}) \\ 
1 - f(x_{\phi'}) + f(y_{\psi'}) & \text{otherwise} 
\end{cases} \\
&= \begin{cases} 
1 & \text{if } f|_{\mathcal{X}}(x_{\phi'}) \leq f|_{\mathcal{X}}(y_{\psi'}) \\ 
1 - f|_{\mathcal{X}}(x_{\phi'}) + f|_{\mathcal{X}}(y_{\psi'}) & \text{otherwise} 
\end{cases} \\
&= f|_{\mathcal{X}}(x_{\phi'}) \lukimplies f|_{\mathcal{X}}(y_{\psi'})
\end{aligned}
\end{equation*}
the third equality follows form our assumptions, the fourth one form induction hypothesis, and the last one from definition of {\luk}.

If $f(x_{\psi'}) < f(x_{\phi'})$, then in this case $f(x) = 1$ (by \ref{form2}), and hence we have:
\begin{equation*}
\begin{aligned}
f(x_\phi) \\
&= 1 - f(x_{\phi'}) + f(y_{\psi'}) \\
&= \begin{cases} 
1 & \text{if } f(x_{\phi'}) \leq f(y_{\psi'}) \\ 
1 - f(x_{\phi'}) + f(y_{\psi'}) & \text{otherwise} 
\end{cases} \\
&= \begin{cases} 
1 & \text{if } f|_{\mathcal{X}}(x_{\phi'}) \leq f|_{\mathcal{X}}(y_{\psi'}) \\ 
1 - f|_{\mathcal{X}}(x_{\phi'}) + f|_{\mathcal{X}}(y_{\psi'}) & \text{otherwise} 
\end{cases} \\
&= f|_{\mathcal{X}}(x_{\phi'}) \lukimplies f|_{\mathcal{X}}(y_{\psi'})
\end{aligned}
\end{equation*}
The first equality holds since \ref{form3}, the second is due to our assumption, the third follows from inductive hypothesis, and the last one due to the definition of $\lukimplies$.
If $f(x_{\psi'}) = f(x_{\phi'})$, note that $f(x) = 1$ or $f(x) = 0$, but in both cases $f(x_\phi) = 1$ and then we reason as the first case.

Case $\phi = \productnot \phi'$: In this case, since $\mathcal{F}(P_{\psi})$, we have that the following equations hold:
\begin{enumerate}
    \item \label{form4}\label{form4} $f(x_{\phi'}) \leq 1 - f(x)$,
    \item \label{form5} $f(x') * (f(x)+f(x_{\phi'})) \geq 1$,
    \item \label{form6} $f(x_\phi) = f(x)$.
\end{enumerate}
The proof proceeds by cases. If $f(x_{\phi}')=0$, then $f(x)=0$ or $f(x)=1$ both satisfies \ref{form4}, but if 
$f(x)=0$ then inequation \ref{form5} does not hold, hence $f(x)=1$ (for which this equation is feasible); 
thus $f(x_\phi) = f(x) = 1 = \productnot 0 = \productnot f(x_{\phi'}) = \productnot f|_{\mathcal{X}}(\phi')$, the last equality holds by induction, which proves this case. If $f(x_{\phi'}) > 0$, 
then by inequation \ref{form4} and taking into account that $x$ is in $\{0,1\}$ we have that $f(x)=0$, thus:
$f(x_\phi) = f(x) = 0 =  \productnot f(x_{\phi'}) = \productnot f|_{\mathcal{X}}(\phi')$, the last equation follows by induction.

Case $\phi = \phi' \productimplies \psi'$: Due to the definition of $\mathcal{F}(P_{\psi})$ the following equations hold:
\begin{enumerate}
    \item \label{form7} $f(x) \geq f(x_{\phi'}) - f(x_{\psi'})$,
    \item \label{form8} $(1-f(x)) \geq f(x_{\psi'}) - f(x_{\phi'})$, 
    \item \label{form8a} $f(x_{\phi}) \geq 1 - f(x)$, 
    \item  \label{form9} $f(x') \cdot (f(x_{\phi'}) - f(x_{\psi'})) \geq f(x)$, 
    \item  \label{form10} $f(x_{\psi'}) - (1-f(x)) \leq f(x_{\phi}) \cdot f(x_{\phi'})$, 
    \item \label{form11} $f(x_{\phi}) \cdot f(x_{\phi'}) \leq f(x_{\psi'}) + (1-f(x))$.
\end{enumerate} 
We proceed by cases. If $f(x_{\phi'}) > f(x_{\psi'})$, then by \ref{form7} we get that $f(x)=1$, then
by inequation \ref{form9}, $f(x')=1$ and so by \ref{form10} and \ref{form11} we get 
$f(x_{\psi'}) =  f(x) \cdot f(x_{\phi'})$, and since $f(x_{\phi'})>0$, we obtain: 
$f(x) = \frac{f(x_{\psi'})}{f(x_{\phi'})} = f(x_{\psi'}) \productimplies f(x_{\phi'}) 
= f|_{\mathcal{X}}(x_{\psi'}) \productimplies f_{\mathcal{X}}(x_{\phi'})$. The last equation follows by induction.
    If $f(x_{\phi'}) < f(x_{\psi'})$, then $f(x)=0$ by equations \ref{form7} and \ref{form8}; thus by inequation
\ref{form8a} we have $f(x_\phi) = 1 = f(x_{\phi'}) \productimplies f(x{\psi'}) = f|_{\mathcal{x}}(x_{\phi'}) \productimplies f|_{\mathcal{X}}(x{\psi'})$. Finally, if $f(x_{\phi'}) = f(x_{\psi'})$ then
either $f(x)=0$ or $f(x)=1$ satisfies equations \ref{form7} and \ref{form8}, but by inequation \ref{form9} $f(x)=0$, and thus
$f(x_\phi) = 1 = f(x_{\phi'}) \productimplies f(x{\psi'}) = f|_{mathcal{x}}(x_{\phi'}) \productimplies f|_{\mathcal{X}}(x{\psi'})$.

Case $\phi = \phi' \godelor \psi'$. Since $f \in \mathcal{F}(P_{\psi})$ we have that the followign equations hold:
\begin{enumerate}
    \item \label{form12} $f(x) \geq f(x_{\phi'}) - f(x_{\psi'})$,
    \item \label{form13} $1-f(x) \geq  f(x_{\psi'}) - f(x_{\phi'})$,
    \item \label{form14} $f(x_\phi) = f(x) \cdot f(x_{\phi'}) + (1-f(x)) \cdot f(x_{\psi})$
\end{enumerate} 
The proof is by cases. If $f(x_{\phi'} > x_{\psi})$ then $f(x)=1$, and therefore:
$f(x_\phi) = f(x_{\phi'}) = f(x_{\psi}) \godelor f(x_{\phi'}) = f|_{\mathcal{X}}(x_{\psi}) \godelor f|_{\mathcal{X}}(x_{\phi'})$. Case
$f(x_{\phi'} \leq x_{\psi})$ is similar.
 
The cases $\phi = \phi' \productor \psi'$ and $\phi = \phi' \productand \psi'$ are direct by induction. 
The case $\phi = \phi' \godeland \psi'$ is simlar to the case $\phi = \phi' \godelor \psi'$.


\end{proof}
\end{lemma}
The next lemma states that any valuation of fuzzy variables over the variables of a  formula $\phi$ can be extended to a function belonging to the feasible set of $P_{\phi}$. Note that here we work on fuzzy formulas (no clauses), which can always be assigned a value. Intuitively, this lemma states that the systems of equations constructed by Algorithm~\ref{alg:mainsat} are not overly restrictive.
\begin{lemma}\label{lemma:SAT-to-MINLP} Let $\phi$ be a fuzzy formula and let $\mathcal{X} = \mathit{Var}(\phi)$, then:
\[
   \forall v \in [0,1]^{\mathcal{X}} : \exists f \in \mathcal{F}(P_\phi) : v = f|_{\mathcal{X}}.
\]
\begin{proof} The proof is by induction on $\phi$.\\
\noindent \textbf{Base Case:} If $\phi = x$ for $x \in \mathcal{X}$, then we just define $f(x) = v(x)$, and similarly if $\phi = k$, for some constant $k$, which proves the result.
\\
\noindent \textbf{Inductive Case:} We consider the possible cases for each logic:

Case $\phi = \neg  \phi'$: It is direct to see that 
$\mathit{Var}(\phi') = \mathcal{X}$, since $\phi'$ is a subformula of $\phi$ and $\phi$ does not have additional variables. Then by induction we have $v = f'|_{\mathcal{X}}$ for some function 
$f' \in \mathcal{F}(P_{\phi'})$. We define a new function $f$ such that
$f(x) = f'(x)$ for $x \in \mathit{Var}(P_{\phi'})$ and $f(x_\phi) = 1 - f(x_{\phi'})$, and so
$f \in \mathcal{F}(P_{\phi})$ since  it satisfies all the equations of $P_\phi$ and also
$f|_{\mathcal{X}} =  f'|_{\mathcal{X}}  = v$.

Case $\phi = \phi' \lukimplies \psi'$: 
Let $\mathcal{X}' = \mathit{Var}(\phi')$ and $\mathcal{X}'' = \mathit{Var}(\psi')$, thus we have:
$\mathcal{X} = \mathcal{X}' \cup \mathcal{X}''$. Furthermore, by induction we have functions $f' \in \mathcal{F}(P_{\phi'})$ and 
$f'' \in \mathcal{F}(P_{\psi'})$ such that $v|_{\mathcal{X}'} = f'|_{\mathcal{X}'}$ and 
$v|_{\mathcal{X}''} = f'|_{\mathcal{X}''}$. We define $f$ over $\mathcal{X}$ as follows:
\[
f(y) = \begin{cases} 
f'(y)  & \text{if } y \in \mathcal{X}' \\ 
f''(y) & \text{if } y \in \mathcal{X}'' \setminus \mathcal{X}' \\
1       & \text{if } y=x \text{ and } f(x_{\phi'}) > f(x_{\psi'}) \\
0       & \text{if } y=x \text{ and } f(x_{\phi'}) \leq f(x_{\psi'}) \\
1 - f(x_{\phi'}) + f(x_{\psi'}) & \text{if } y = x_\phi  \text{ and } f(x_{\phi'}) > f(x_{\psi'}) \\
1  & \text{if } y = x_\phi  \text{ and } f(x_{\phi'}) \leq f(x_{\psi'}) \\
\end{cases} \\
\]
By definition this function satisfies all the inequations in $P_{\phi'}$ and 
$P_{\psi'}$, let us prove that it satisfies the remaining equations, i.e.:
\begin{enumerate}
    \item \label{formluk1} $f(x) \geq f(x_{\phi'}) - f(x_{\psi'})$,
    \item \label{formluk2} $1-f(x) \geq f(x_{\psi'}) - f(x_{\phi'})$,
    \item \label{formluk3} $f(x_\phi) = (1-f(x))  + f(x) \cdot(1 - f(x_{\phi'}) + f(x_{\psi'}))$.
\end{enumerate}
If $f(x_{\phi'}) > f(x_{\psi'})$ then $f(x)=1$ which satisfies \ref{formluk1} and
$\ref{formluk2}$ since $f(x_{\phi'}) - f(x_{\psi'}) > 0$, furthermore
$f(x_\phi) = 1 - f(x_{\phi'}) + f(x_{\psi'})$ which satisfies \ref{formluk3}.
If $f(x_{\phi'}) \leq f(x_{\psi'})$ then $f(x)=0$  which satisfies \ref{formluk1} and
$\ref{formluk2}$ and furthermore $f(x_\phi) = 1$, thus:
$(1-f(x))  + f(x) \cdot(1 - f(x_{\phi'}) + f(x_{\psi'})) = 1 - f(x_{\phi'}) + f(x_{\psi'}) = f(x_\phi)$, making \ref{formluk3} true.
Furthermore, by definition $f(x) = v(x)$ for $x \in \mathcal{X}' \cup \mathcal{X}''$ and so $f|_{\mathcal{X}} = v$. 

Case $\phi = \productnot \phi'$:  As the case of {\luk} negation, we have that $\mathit{Var}(\phi') = \mathcal{X}$.  By induction we have $v = f'|_{\mathcal{X}}$ for some function 
$f' \in \mathcal{F}(P_{\phi'})$. We define a new function $f$ as follows:
\[
f(y) = \begin{cases} 
f'(y)  & \text{if } y \in \mathit{Var}(P_{\phi'}) \\ 
1       & \text{if } y=x_\phi \text{ and } f(x_{\phi'}) = 0 \\
0       & \text{if } y=x_\phi \text{ and } f(x_{\phi'}) > 0 \\
0       & \text{if } y=x' \text{ and } f(x_{\phi'}) > 0 \\
1       & \text{if } f(x_{\phi'}) = 0 \text{ and } y=x' \\
\frac{1}{f(x_{\phi'})}        & \text{if }f(x_{\phi'}) = 1 \text{ and } y=x'
\end{cases} \\
\]
By induction, this function satisfies the inequations in  $P_{\phi'}$. Let us check it satisfies the additional equations in $P_\phi$, i.e.:
\begin{enumerate}
    \item \label{form15}\label{form4} $f(x_{\phi'}) \leq 1 - f(x_\phi)$,
    \item \label{form16} $f(x') * (f(x_\phi)+f(x_{\phi'})) \geq 1$,
\end{enumerate}
If $f(x_{\phi'})=0$, then $f(x_\phi) = 1$ and equation \ref{form15} because both terms are evaluated to $0$.
Equation \ref{form16} holds since $f(x)+f(x_{\phi'}) > 0$ and  $f(x')=1$ makes the equation true.
If $f(x_{\phi'})>0$ then then $f(x_\phi) = 0$, which makes inequality \ref{form15} true, and
since $f(x') = \frac{1}{f(x_{\phi'})}$ and $f(x_{\phi'})>0$ we have that \ref{form16} trivially holds. Furthermore, by induction we have $f|_{\mathcal{X}} = v$.

Case $\phi = \phi' \productimplies \psi'$: As in the case of {\luk},  we set  $\mathcal{X}' = \mathit{Var}(\phi')$ and $\mathcal{X}'' = \mathit{Var}(\psi')$. Furthermore,  we have 
$X = X' \cup X''$. By induction we have functions $f' \in \mathcal{F}(\phi')$ and 
$f'' \in \mathcal{\psi'}$ such that $v|_{\mathcal{X}'} = f'|_{\mathcal{X}'}$ and 
$v|_{\mathcal{X}''} = f'|_{\mathcal{X}''}$. We define $f$ over $\mathcal{X}$ as follows:
\[
f(y) = \begin{cases} 
f'(y)  & \text{if } y \in \mathcal{X}' \\ 
f''(y) & \text{if } y \in \mathcal{X}'' \\
1       & \text{if } y=x_\phi \text{ and } f(x_{\phi'}) \leq f(x_{\psi'}) \\
\frac{f(x_{\psi'})}{f(x_{\phi'})}       & \text{if } y=x_\phi \text{ and } f(x_{\phi'}) > f(x_{\psi'}) \\
1  & \text{if }  y = x \text{ and }  f(x_{\phi'}) \leq f(x_{\psi'}) \\
0  & \text{if } y = x \text{ and }  f(x_{\phi'}) > f(x_{\psi'}) \\
\frac{1}{f(x_{\phi'}) - f(x_{\psi'})}  & \text{if }  y = x' \text{ and }  f(x_{\phi'}) > f(x_{\psi'}) \\
0  & \text{if }  y = x' \text{ and }  f(x_{\phi'}) \leq f(x_{\psi'}) \\
\end{cases} \\
\]
Now, we can verify that this function satisfies the corresponding equations:
\begin{enumerate}
    \item \label{form17} $f(x) \geq f(x_{\phi'}) - f(x_{\psi'})$,
    \item \label{form18} $(1-f(x)) \geq f(x_{\psi'}) - f(x_{\phi'})$, 
    \item \label{form19} $f(x_{\phi}) \geq 1 - f(x)$, 
    \item  \label{form9} $f(x') \cdot (f(x_{\phi'}) - f(x_{\psi'})) \geq f(x)$, 
    \item  \label{form10} $f(x_{\psi'}) - (1-f(x)) \leq f(x_{\phi}) \cdot f(x_{\phi'})$, 
    \item \label{form11} $f(x_{\phi}) \cdot f(x_{\phi'}) \leq f(x_{\psi'}) + (1-f(x))$.
\end{enumerate} 
If $f(x_{\phi'}) \leq f(x_{\psi'})$ then we have $0 \leq f(x_{\psi'}) -  f(x_{\phi'})$ and equations \ref{form17} and \ref{form18} hold.
Equation \ref{form19} holds since $1 - f(x) = 0$. Equation \ref{form9} hold since $f(x') = 0$ and 
$f((x)=0$. Equation~\ref{form10} and ~\ref{form11} hold since $f(x)=0$ implies that they 
are equivalent to $f(x_{\psi'}) - 1 \leq f(x_{\phi}) \cdot f(x_{\phi'}) \leq f(x_{\psi'}) + 1$
and this is true because all variables take values in $[0,1]$. If $f(x_{\phi'}) > f(x_{\psi'})$ then
$f(x)=1$ and equations \ref{form17} and \ref{form18} since $f(x_{\psi'}) - f(x_{\phi'}) < 0$. 
Inequation \ref{form19} holds since all the variables are in $[0,1]$. Since $f(x)=1$ inequations
\ref{form10} and \ref{form11} are equivalent to $f(x_{phi'}) \leq f(x_\phi) \cdot  f(x_\psi') \leq f(x_{\psi'})$, which holds since $f(x_\phi) = \frac{f(x_{\psi'})}{f(x_{\phi'})}$.
\end{proof}
\end{lemma}

The correctness of this algorithm boils down to proving the following theorems. The first theorem establishes a strong correspondence between SAT valuations and the feasible set of the MINLP problem constructed by Algorithm~\ref{alg:mainsat}. 
\begin{theorem}\label{th:iso} Given a (finite) set of fuzzy clauses (in {\luk}, G\"odel, or product logic) $\Phi$ with $\mathit{Var}(\Phi) = \mathcal{X}$, and let $P_\Phi$ be the MINLP problem constructed in line~~\ref{alg:minlp-problem} of Algorithm~\ref{alg:mainsat}. We have that: 
$\{ f|_{\mathcal{X}} \mid f \in \mathcal{F}(P_\Phi) \} = \{v \in [0,1]^\mathcal{X} \mid v \vDash_\infty \Phi \}$.
\begin{proof}
We prove the theorem for one clause; it is straightforward to extend this proof to several clauses. Let $ \ell \leq \phi \leq u$ be a fuzzy clause. Without loss of generality we assume $\mathcal{X} = \mathit{Var}(\phi)$.
 Let $v \in [0,1]^\mathcal{X}$ be a valuation such that $v \vDash_\infty  \ell \leq \phi \leq u$, that is,
$\ell \leq v(\phi) \leq u$, by Lemma \ref{lemma:SAT-to-MINLP} there is a $f \in \mathcal{F}(P_\phi)$ such that $f|_{\mathcal{X}} = v$, and so by Lemma \ref{lemma:MINLP-to-SAT} 
we get $f(x_\phi) = v(\phi)$, then $\ell \leq f(x_\phi) \leq u$ which proves $\{v \in [0,1]^\mathcal{X} \mid v \vDash_\infty \ell \leq \phi \leq u\} \subseteq 
\{ f|_{\mathcal{X}}  \mid f \in \mathcal{O}(P_\Phi) \}$. Now, let $f|_{\mathcal{X}} \in \{ f|_{\mathcal{X}}  \mid f \in \mathcal{O}(P_\Phi) \}$, then 
$\ell \leq f(x_\phi) \leq u$, and so by Lemma \ref{lemma:MINLP-to-SAT} we get $\ell \leq f|_{\mathcal{X}}(\phi) \leq u$ which implies that
$f|_{\mathcal{X}} \in \{v \in [0,1]^\mathcal{X} \mid v \vDash_\infty \ell \leq \phi \leq u\}$ and therefore:
$\{ f|_{\mathcal{X}}  \mid f \in \mathcal{F}(P_\Phi) \} \subseteq \{v \in [0,1]^\mathcal{X} \mid v \vDash_\infty \ell \leq \phi \leq u\}$ which proves the theorem.
\end{proof}
\end{theorem} 

\begin{theorem}\label{theorem:correctness} A set of fuzzy clauses (in {\luk}, G\"odel, or product logic)  $\Phi$ is $SAT_\infty$ iff $\textsc{SAT}(\Phi)$ returns \say{SAT}.
\begin{proof}
The result follows from Theorem~\ref{th:iso}. If Alg.~\ref{alg:mainsat} returns \say{SAT} then there is a $f \in \mathcal{F}(P_{\Phi})$ such that
$\ell \leq f(x_\phi) \leq u$, but then by Theorem \ref{th:iso} we have a valuation $v$ such that $\ell \leq v(\phi) \leq u$ and the formula is SAT.
If there is a valuation $v$ such that $\ell \leq v(\phi) \leq u$, then by Lemma~\ref{lemma:SAT-to-MINLP} we have that there is an assignment $f \in \mathcal{F}(P_\Phi)$ such that $f|_{\mathcal{X}} = v$ and so $\ell \leq f(x_\phi) \leq u$, then Alg.~\ref{alg:mainsat} returns \say{SAT} given that the solver used is complete.
\end{proof}
\end{theorem}
Note that, in practice, the completeness of the method depends on the solver used. Solvers like \textsf{Gurobi}~\cite{gurobi} or \textsf{SCIP}~\cite{BolusaniEtal2024OO} use finite-precision arithmetic; thus, it may be that the solver cannot find certain solutions. On the other hand, for solvers that use infinite-precision arithmetic, e.g., \textsf{Yices}, the completeness of the method is guaranteed.

\section{Logical and Numerical Optimizations} \label{sec:opt}
In this section, we consider some numerical and logical optimizations to Algorithm~\ref{alg:to-minlp}. First, we observe that the arithmetical representations of product negation and product implication are ones of the most \emph{expensive} in numerical terms since it involves solving inequality $x' \cdot (x_{\phi} + x_{\phi'}) \geq 1$ ($\dagger$), in the case of $\productnot$, and 
$x' \cdot (x_{\phi'} - x_{\psi'}) \geq x$ ($\ddagger$), in the case of $\productimplies$; which are non-convex quadratic expressions,  typically hard to solve for optimization tools. This can be observed from the experiments reported in Section~\ref{sec:experiments}. We can avoid the quadratic formula ($\dagger$)
noting that  this formula is used to prevent $x_{\phi}$ and $x_{\phi'}$ from being $0$ at the same time.
The following reformulation (standard in optimization): $x_{\phi} + x_{\phi'} >= \epsilon$ can be used to formulate this restriction, avoiding quadratic constraints, where $\epsilon$ is the smallest number representable by the optimizer.

Furthermore, a main technique we propose for optimization is pushing negations inside formulas, which also allows us to remove implications and thus avoid ($\ddagger$) when possible. It is worth observing that the law of double negation does not hold for $\productnot$ (or $\godelnot$). However, we can see that double product negation is equivalent to the following function:
\[
\ominus x = \begin{cases}
                                0 & \text{if } x = 0\\
                                1 & \text{otherwise }
                            \end{cases}
\]
One way to compute this operator is with the following system:
\[
x \leq z, z \geq x * \epsilon
\]
where $z$ is a Boolean variable having the value of $\productnot \productnot x$, and $\epsilon$ is the smallest number representable by the selected solver. Thus, we push negations inside of formulas and replace double negations with this operator in the resulting formula.

Let us consider the following properties of $\productnot$ (or equivalently $\godelnot$):
\begin{enumerate}
    \item $\productnot (x \productand y) = (\productnot x) \productor (\productnot y)$,
    \item $\productnot (x \productor y) = (\productnot x) \productand (\productnot y)$,
    \item \label{rule-pimplication} $\productnot (x \productimplies y) = (\productnot \productnot x) \productand (\productnot y)$,
    \item $\godelnot (x \godeland y) = (\godelnot x) \godelor (\godelnot y)$,
    \item $\godelnot (x \godelor y) = (\godelnot x) \godeland (\godelnot y)$,
     \item \label{triple-product-negation} $\productnot \productnot \productnot x = \productnot x$.
    \item \label{double-product-negation} $\productnot \productnot x = \ominus x$.
\end{enumerate}
Thus, using these properties, we can push the product negation inside until we reach variables. Furthermore, rule \ref{rule-pimplication} allows us to remove product implications when possible.
Rule \ref{triple-product-negation} makes it possible to reduce product negations. At the end of this process, we obtain formulas in which there are only single or double negations over propositional variables. Finally, we can use rule \label{double-product-negation} to transform double product negations into $\ominus$, which is easier to deal with for the arithmetical solvers. Algorithm~\ref{alg:sat-opt} shows the optimization procedure.
\begin{algorithm}[!h]
\small
\SetKwInOut{Input}{Input}
\SetKwInOut{Output}{Output}
\SetKwProg{Fn}{Function}{}{end}

\Input{A fuzzy formula $\phi$}
\Output{An optimized fuzzy formula $\phi'$}

\caption{Algorithm for Formula Optimization}\label{alg:sat-opt}

\Fn{\textsc{Opt}($\phi$)}{
    \lIf{$\phi = x$}{
        \KwRet $x$
    }
    
    \If{$\phi = \neg \phi'$ where $\neg \in \{\godelnot, \productnot \}$}{
        \lIf{$\phi' = x$}{
            \KwRet $\phi$
        }
        \lElseIf{$\phi' = \neg\neg \phi''$}{
            \KwRet \textsc{Opt}($\neg \phi''$)
        }
        \lElseIf{$\phi' = \neg x$ for $x\in \mathit{AP}$}{
            \KwRet $\neg \neg x$
        }
        \lElseIf{$\phi' = \neg \phi''$}{
            $\psi \gets \textsc{Opt}(\neg \phi'')$;
            \KwRet \textsc{Opt}($\neg \psi$)}
        \lElseIf{$\phi' = \phi'' \oplus \phi'''$}{
            \KwRet \textsc{Opt}($\neg \phi''$) $\overline{\oplus}$ \textsc{Opt}($\neg \phi'''$)
        }
    }
    
    \lIf{$\phi = \phi' \oplus \phi''$ and $\oplus \in \{\godeland,\godelor,\productand,\productor \}$}{
        \KwRet \textsc{Opt}($\phi'$) $\oplus$ \textsc{Opt}($\phi''$)
    }
}
\end{algorithm}
Furthermore, instead of  solving the equation $x' \cdot (x_{\phi} + x_{\phi'}) \geq 1$ we solve the equation $(x_{\phi} + x_{\phi'}) \geq \epsilon$, being $\epsilon$ the smallest number representable in solver's precision. 

\section{Related Work} \label{sec:related}
It is worth comparing Algorithm~\ref{alg:mainsat} with related approaches.
In \cite{Haehnle1994}, a reduction of many-valued logics to MILP  is introduced. In this work, only logics expressible in MILP are discussed, i.e., the product logic is excluded from this approach.  Furthermore, to reduce $\phi' \lukimplies \psi' \leq u$ the following constraints are used: $x_{\phi'} \geq i_1$, $x_{\psi'} \leq i_2$, $u + y= 1 - i_1 +i_2$, $y \leq u$, $i_1 \leq 1-y$, $y \leq i_2$, where $y \in \{0,1\}$. The rule is different for $\phi' \lukimplies \psi' \geq l$. Note that in contrast to this approach, our algorithm leverages multiplication to use the same equations for $\leq$ and $\geq$.  A similar procedure is used in \cite{Hajek1998} to reduce {\luk} logic to MILP problems. \textsf{fuzzyDL}~\cite{Bobillo2016} is a description logic reasoner that supports {\luk} fuzzy logic reasoning. SAT problems can be codified with knowledge databases, and SAT queries can be performed with this tool. \textsf{fuzzyDL} uses a tableau procedure together with MILP solvers (\text{Gurobi} or \textsf{CBC} \cite{forrest2005cbc}) to solve queries over knowledge databases. As stated in \cite{Vidal2016}, \textsf{fuzzyDL} is not designed to support other logics like product logics.
It is worth stressing that in previous works only {\L}ukasiewicz and similar logics are reduced to MILP; as far as we are aware, reductions of product logic to arithmetic constraints  are not discussed in related literature. 
Furthermore, it is worth noting that Theorem~\ref{th:iso} establishes that our translation to MINLP problems preserves solutions. A similar property is not proven for the translations given in \cite{Hajek1998,Haehnle1994}. This could be relevant when considering extensions of \textsf{SATFuL}, for instance, to inspect the set of solutions. In \cite{Vidal2016}, a SAT solver (\textsf{MNiBLoS}) able to deal with several fuzzy logics is presented; \textsf{MNiBLoS} translates fuzzy constraints to SMT theories over arithmetic theories. Particularly, \textsf{MNiBLoS} translates product logic to negative integer arithmetic. However, this approach is based on an incomplete
transformation of nonlinear arithmetic problems into negative arithmetic. As a result,
unsatisfiable clauses may be incorrectly classified as satisfiable by \textsf{MNiBLoS}, a pathological example is the formula: 
$\{0.75 \leq \productnot (x_1 \godelimplies x_2) \godelimplies x_3 \leq 0.75, 0 \leq x_3 \leq 0.5 \}$.
On the other hand, in \cite{DBLP:journals/sncs/Uhliarik22} a solver is presented for product logic extended with Monteiro-Baaz modal connective. However, this solver targets 1-SAT over these formulas and does not consider clauses with constants, so it cannot be used to solve $k$-SAT over fuzzy clauses as done here.


%
\section{Experimental Results} \label{sec:experiments}
We conducted an initial evaluation of \textsf{SATFuL} and compared its performance with related tools: 
\textsf{fuzzyDL}, a well-known description logic reasoner supporting fuzzy logic and fuzzy rough set reasoning~\cite{Bobillo2016}, which is able to solve {\luk} and G\"odel logics using a logical preprocessing and linear programming solvers, we set this tool to work with \textsf{Python-MIP}\footnote{https://github.com/coin-or/python-mip} as well as \textsf{Gurobi} \cite{gurobi}, to avoid any bias introduced by the arithmetic solvers. We have also included the SMT solver \textsf{Yices}, which provides support for non-linear arithmetic; in this case, conditional operators can be straightforwardly encoded since it also provides support for \emph{if-then-else} constructions, while the other operators are codified as in Algorithm~\ref{alg:to-minlp}. On the other hand, for product logic we compare also with  
\textsf{MNiBLoS} \cite{Vidal2016}, a product logic SAT solver that uses the SMT solver \textsf{Z3}~\cite{DeMoura08}. All experiments were performed on a MacBook M2 with 16 GB of RAM. \textsf{SATFuL} is an open-source tool entirely programmed with \textsf{Python} available in \url{https://github.com/pablofcastro/satful}.
%
%
%
%
\paragraph{{\luk} Logic}
First, we compare \textsf{SATFuL} and \textsf{fuzzyDL} using a benchmark of formulas proposed in \cite{DBLP:journals/jar/SchockaertJV12}. This benchmark consists of 200 randomly generated formulas generated using the following process. Given a set $\mathcal{V} = \{v_1,\dots,v_n\}$ of propositional variables, a recursive procedure $\mathit{F}$ is defined such that $\mathit{F}(m)$ returns a {\luk} formula with $m$ variable occurrences. For the base case, we define:
\[
\text{F}(1) = 
\begin{cases} 
v_i & \text{with probability } 0.5, \\
\neg v_i & \text{with probability } 0.5.
\end{cases}
\]
For a randomly picked variable $v_i \in \mathcal{V}$ with a uniform distribution. The recursive case is defined as follows: 
\[
\mathit{F}(p) = 
\begin{cases} 
\mathit{F}(p_1) \lukand \mathit{F}(p_2) & \text{with probability } 0.5, \\
\luknot (\mathit{F}(p_1) \lukand \mathit{F}(p_2)) & \text{with probability } 0.5, 
\end{cases}
\]
where $p_1$ is an integer randomly chosen from interval $[1,p-1]$ and $p_2 = p - p_1$. Note that standard procedures for generating SAT instances of propositional logic \cite{Selman96} cannot be used in the fuzzy setting since not all {\luk} formulas can be converted to CNF. Furthermore, {\luk} clauses in conjunctive normal form with only lower bounds can be translated to linear programming problems, and therefore solved in (pseudo-)polynomial time. 

Furthermore, to obtain interesting problems, the lower and upper bounds for the clauses must be carefully selected. In this case, the procedure is the following. Lower and upper bounds are picked from the set $T = \{0,0.25,0.50,0.75,1\}$ as follows. First, $k$ formulas are generated with the procedure described above. For each formula $\alpha$ (in this set), two random interpretations $\chi$ and $\chi'$ over $V$ are generated, and then the largest value $\ell$ from $T$ such that $\ell \leq \chi(\alpha)$ is selected as lower bound, and the largest value such that $u \leq \chi'(\alpha)$, then the clause $\ell \leq \alpha \leq u$ is added to the set of clauses. Using this procedure, $200$ test sets with $10,20,30$ and $40$ variables and $100$ fuzzy clauses were generated.  It is clear that the fewer variables, the more constrained the problems become. 

\begin{figure}[!htb]
\centering     
   \includegraphics[scale=0.5]{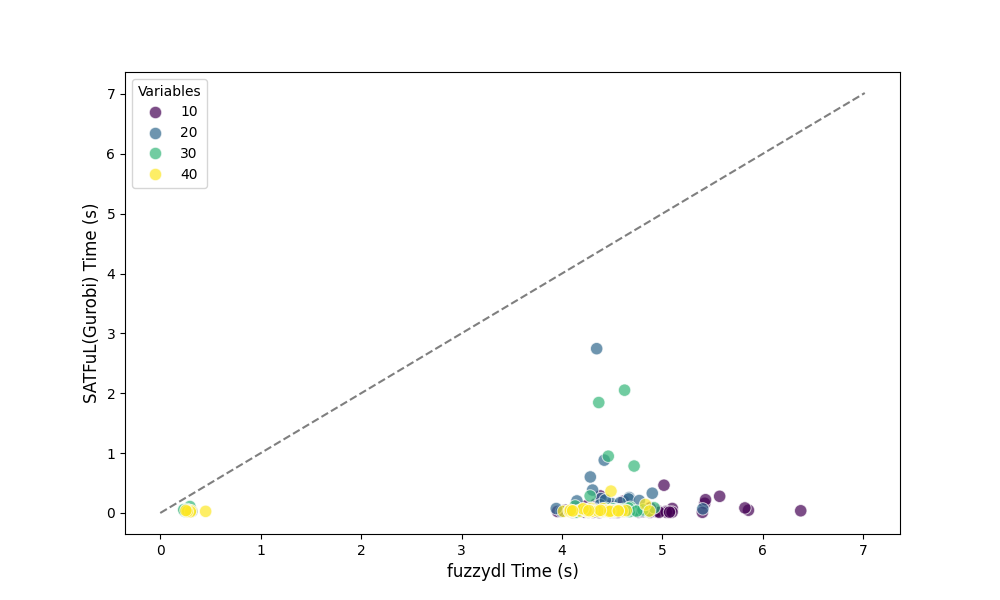}
   \caption{\textsf{SATFuL} vs \textsf{fuzzyDL} over number of variables}
   \label{fig:satful-gurobi-vs-fuzzysat-vars}
\end{figure}

\begin{figure}[!htb]
\centering     
   \includegraphics[scale=0.5]{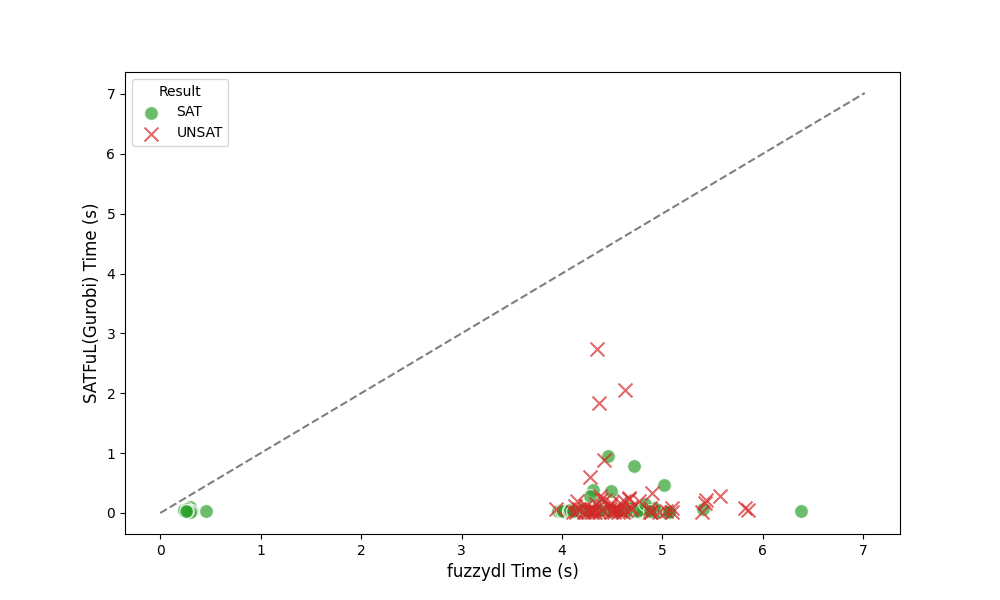}
   \caption{\textsf{SATFuL} vs \textsf{fuzzyDL} for SAT/UNSAT results}
   \label{fig:satful-gurobi-vs-fuzzysat-sat}
\end{figure}

The performance of \textsf{SATFuL} and \textsf{fuzzyDL} on this set of formulas is shown in Figure \ref{fig:satful-gurobi-vs-fuzzysat-vars} (grouped by number of variables) and Figure \ref{fig:satful-gurobi-vs-fuzzysat-sat} (distinguishing SAT from UNSAT instances). For consistency, both solvers were configured to use \textsf{Gurobi} 10.0.3 for arithmetic solving.
As shown in these figures, \textsf{SATFuL} outperformed \textsf{fuzzyDL} in all cases. The average speedup of \textsf{SATFuL} over \textsf{fuzzyDL} was $99.5\times$, with a maximum speedup reaching $471.54\times$ for several problems with 40 variables. \textsf{SATFuL} achieved a mean solving time of $0.11$ seconds, whereas \textsf{fuzzyDL} showed a mean time of $3.74$ seconds. As illustrated in Figure~\ref{fig:satful-gurobi-vs-fuzzysat-sat}, both SAT and UNSAT problems follow similar performance trends for both solvers. However, \textsf{SATFuL} exhibits slightly higher latency in a small subset of UNSAT problems with 20 and 30 variables. 
The two solvers showed a 100\% coincidence in the results.
\begin{table}[ht]
\centering
\footnotesize 
\setlength{\tabcolsep}{3.5pt} 
\caption{Performance comparison between \textsf{fuzzyDL} ($x$) and \textsf{SATFuL} ($y$).}
\label{tab:performance_complete}
\begin{tabular}{@{}lccccc@{}} 
\toprule
\multirow{2}{*}{\textbf{Vars}} & \multicolumn{2}{c}{\textbf{Time \textsf{fuzzyDL} (s)}} & \multicolumn{2}{c}{\textbf{Time \textsf{SATFuL} (s)}} & \textbf{Speedup} \\ \cmidrule(lr){2-3} \cmidrule(lr){4-5} \cmidrule(l){6-6} 
 & \textit{Mean} & \textit{Max} & \textit{Mean} & \textit{Max} & \textit{Mean} \\ \midrule
10 & 4.022 & 4.466 & 0.069 & 0.464 & 148.69$\times$ \\
20 & 3.824 & 4.510 & 0.169 & 2.745 & 74.50$\times$ \\
30 & 3.479 & 4.410 & 0.161 & 2.052 & 77.06$\times$ \\
40 & 3.651 & 4.532 & 0.046 & 0.365 & 97.74$\times$ \\ \midrule
\end{tabular}
\end{table}
Table~\ref{tab:performance_complete} shows the results, grouping problems by the number of variables. It can be observed that the group of problems with 10 variables appears to be more complex to solve for \textsf{fuzzyDL}, as it contains more UNSAT instances.

\begin{figure}[!htb]
\centering     
   \includegraphics[scale=0.4]{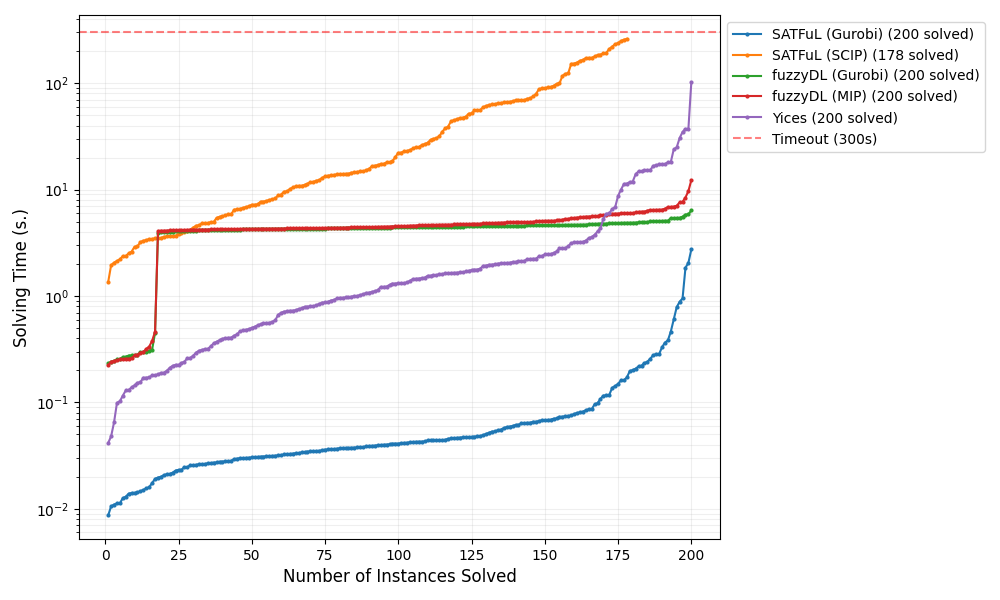}
   \caption{\textsf{SATFuL} vs related tools}
   \label{fig:luk-cactus-plot}
\end{figure}

Furthermore, we can compare the performance of \textsf{SATFuL} with \textsf{Yices}, a modern SMT solver that supports linear arithmetic and conditional operators such as $\max$ and $\min$, thus {\luk} clauses can be solved using it. Also, to check the impact of using \textsf{Gurobi} on the tool's performance, we have also run \textsf{SATSuL} using \textsf{SCIP} \cite{BolusaniEtal2024OO}, an open-source library supporting non-linear arithmetic. Figure~\ref{fig:luk-cactus-plot} shows the cactus plot obtained when all these tools are compared.
Some comments about the results obtained are useful. First, note that \textsf{SATFuL} using \textsf{Gurobi} as an arithmetic solver is the fastest of all options, which includes \textsf{fuzzyDL} using \textsf{Gurobi}. On the other hand, \textsf{fuzzyDL} shows a jump in time due to the preprocessing required to transform the knowledge database into MILP problems. It must be noted that the performance of \textsf{fuzzyDL} using \textsf{Gurobi} and \textsf{MIP} is similar, which indicates that after the preprocessing, the obtained linear problems are simple. \textsf{Yices} behaves well for problems with few variables, but it becomes slower when 20-30 variables are considered, which is the ratio \emph{clause/variables} where the formulas get harder.
Finally, \textsf{SATFuL} using \textsf{SCIP} is slower than the other options; it is not able to solve 22 cases (most of them are problems with 20 variables) before the time-out, which shows that the performance of the backend solver affects Algorithm~\ref{alg:mainsat}.

\paragraph{G\"odel Logic}

In order to evaluate the SAT solvers for G\"odel fuzzy logic, we  adapted the procedure for generating formulas discussed above in the following manner. We consider the following minimal set of operators for  the G\"odel logic: $\{ \godeland, \godelor, \godelnot \}$,  and we modify the recursive case of function $\mathit{F}$  as follows:
\[
\mathit{F}(p) = 
\begin{cases} 
\mathit{F}(p_1) \godelor \mathit{F}(p_2) & \text{with prob. } 0.25, \\
\godelnot (\mathit{F}(p_1) \godelor \mathit{F}(p_2)) & \text{with prob. } 0.25,\\
\mathit{F}(p_1) \godeland \mathit{F}(p_2) & \text{with prob. } 0.25, \\
\godelnot (\mathit{F}(p_1) \godeland \mathit{F}(p_2)) & \text{with prob. } 0.25.
\end{cases}
\]

\begin{figure}[!htb]
\centering     
   \includegraphics[scale=0.4]{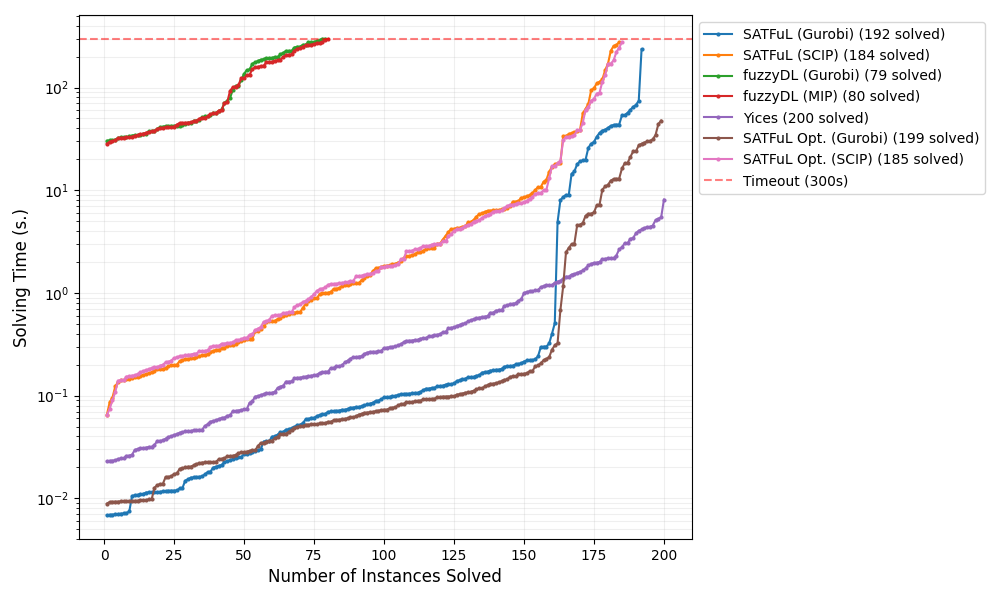}
   \caption{\textsf{SATFuL} vs related tools over G\"odel formulas}
   \label{fig:cactus-godel}
\end{figure}

Using this procedure, we have generated 200 problems, each consisting of 100 clauses with 10, 20, 30, or 40 variables. We use this benchmark to compare the performance of \textsf{SATFuL}, \textsf{Yices}, and \textsf{FuzzyDL}; we also considered \textsf{SATFuL} running the optimized algorithm described in Section~\ref{sec:opt}.
Figure~\ref{fig:cactus-godel} shows a cactus plot describing the results obtained.
As shown in this figure, \textsf{FuzzyDL} is not able to deal with most problems of the benchmark before the timeout. Specifically, it cannot solve any problem with most of 20 variables before the timeout.  G\"odel operators seem to negatively affect the performance of the preprocessing techniques used by \textsf{FuzzyDL}, as it shows the same behavior when using the \textsf{MIP} solver as the \textsf{Gurobi} solver. On the other hand, \textsf{Yices} shows good performance and was able to solve all the problems in less than 300 seconds, this can be due to the native support of $\max$ and $\min$ operators provided by \textsf{Yices}. 

On the other hand, \textsf{SATFuL}  with \textsf{Gurobi} offers a similar performance to \textsf{Yices} for formulas of 10 and 20 variables, but \textsf{Yices}  performance is better for formulas of 30 and 40 variables. Furthermore, as shown in the figure, the optimizations discussed in Section~\ref{sec:opt} improve the tool's performance, but it remains slower than \textsf{Yices} for some formulas with 40 variables.  Table~\ref{tab:solver_performance_3dec} shows a more detailed comparison 
of the median of the running times.
\begin{table}[htbp]
\centering
\caption{Median Execution Time (s) by Number of Variables}
\label{tab:solver_performance_3dec}
\begin{tabular}{c *{7}{w{c}{1.2cm}}}
\toprule
\textbf{\texttt{num\_vars}} & \textbf{\shortstack{SATFuL\\(Gurobi)}} & \textbf{\shortstack{SATFuL\\(Gurobi Opt.)}} & \textbf{\shortstack{SATFuL\\(SCIP)}} & \textbf{\shortstack{SATFuL\\(SCIP Opt.)}} & \textbf{Yices} & \textbf{\shortstack{FuzzyDL\\(Gurobi)}} & \textbf{\shortstack{FuzzyDL\\(MIP)}} \\
\midrule
\textbf{10} & \textbf{0.012} & 0.017          & 0.199  & 0.234  & 0.042          & 42.456  & 42.646  \\
\textbf{20} & 0.064          & \textbf{0.053} & 0.888  & 1.075  & 0.239          & 280.987 & 263.532 \\
\textbf{30} & 0.163          & \textbf{0.125} & 5.027  & 4.808  & 0.619          & 300.000 & 300.000 \\
\textbf{40} & 11.734         & 2.092          & 37.124 & 35.730 & \textbf{1.830} & 300.000 & 300.000 \\
\bottomrule
\end{tabular}
\end{table}
Some observations about this table are useful. \textsf{SATFuL} using \textsf{Gurobi}  outperforms \textsf{SATFuL} with \textsf{SCIP} by 10x to 30x across all the instance sizes. While unoptimized \textsf{SATFuL} is marginally faster than optimized \textsf{SATFuL} when using \textsf{Gurobi} (0.012s vs 0.017s) with $10$ vars; optimization becomes important when  the number of variables increases, yielding a 5.6x speedup at 40 variables (2.09s vs 11.73s). The optimized algorithm also behaves better when \textsf{SCIP} is used, although the difference is much smaller, which can be attributed to the non-linear procedures used by \textsf{SCIP}. 

Regarding \textsf{Yices}, it demonstrates greater scalability than the unoptimized \textsf{SATFuL}, exhibiting a flatter growth curve as the number of variables increases. Although starting slightly behind \textsf{SATFuL} with \textsf{Gurobi} at 10 variables, \textsf{Yices} slows down by only ~3x between 30 and 40 variables (0.62s to 1.83s), whereas unoptimized \textsf{SATFuL} with \textsf{Gurobi} spikes 72x over the same range. In contrast, \textsf{FuzzyDL} suffers a severe bottleneck, running 3 to 4 orders of magnitude slower across all configurations and consistently hitting the 300-second execution timeout starting at 30 variables.

It is interesting to compare optimized \textsf{SATFuL} using \textsf{Gurobi} with \textsf{Yices}, these are the two solvers that behave better for this benchmark.  While optimized SATFuL with Gurobi behaves better in small and medium problems (10–30 variables), \textsf{Yices} takes over at 40 variables due to its  better scaling. For tiny instances (10 variables), optimized \textsf{SATFuL} with \textsf{Gurobi}  is roughly 2.5× faster than \textsf {Yices} (0.017s vs. 0.042s), and it maintains a 5x lead at 30 variables (0.125s vs. 0.619s). This makes optimized \textsf{SATFuL} using \textsf{Gurobi} a good choice when low latency on smaller instances is a priority.

However, for 30-40 variables, optimized \textsf{SATFuL} with \textsf{Gurobi} experiences a ~17x increase in execution time (0.125s to 2.092s), whereas \textsf{Yices} grows by only ~3× (0.619s to 1.830s).

\begin{figure}[!htb]
\centering     
   \includegraphics[scale=0.5]{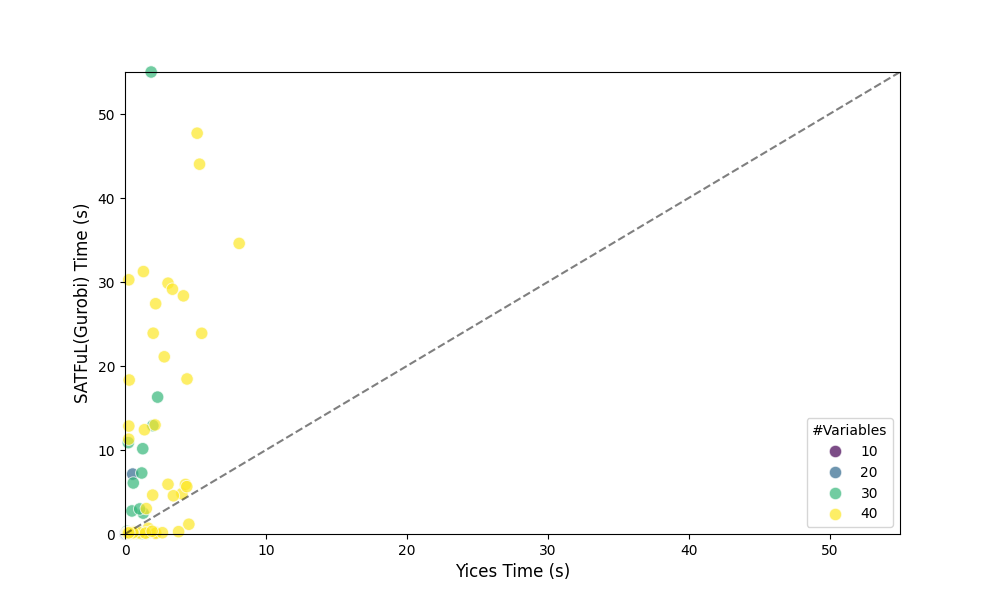}
   \caption{\textsf{SATFuL} vs Yices over G\"odel formulas wrt number of variables}
   \label{fig:satful-gurobi-vs-yices-vars-godel}
\end{figure}

\begin{figure}[!htb]
\centering     
   \includegraphics[scale=0.5]{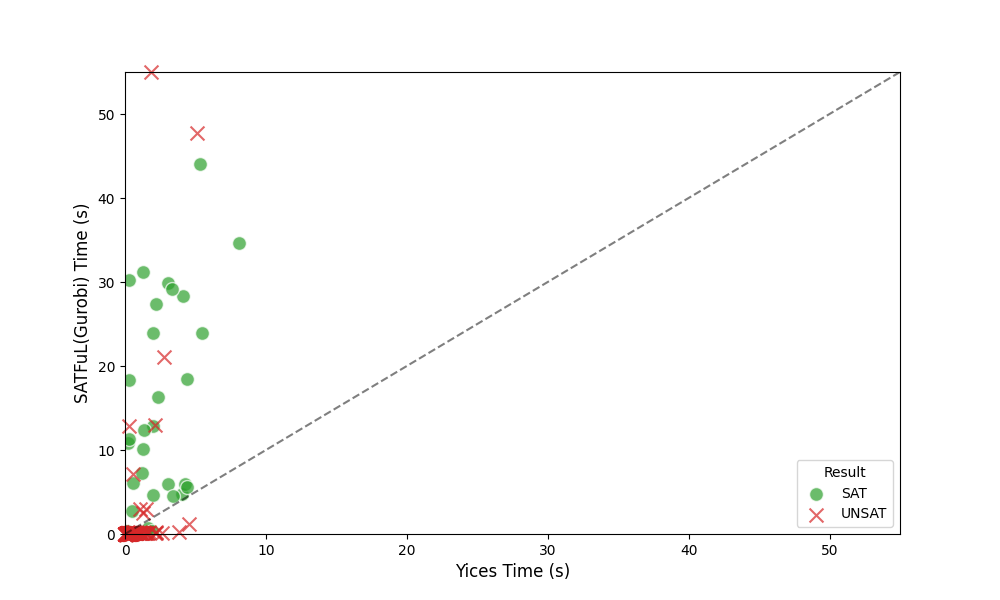}
   \caption{\textsf{SATFuL} vs Yices over G\"odel formulas wrt SAT/UNSAT}
   \label{fig:satful-gurobi-vs-yices-sat-godel}
\end{figure}

A more detailed comparison between \textsf{Yices} and optimized \textsf{SATFuL} using \textsf{Gurobi} can be obtained by observing Figure~\ref{fig:satful-gurobi-vs-yices-vars-godel} and Figure~\ref{fig:satful-gurobi-vs-yices-sat-godel}, which compare the performance of both solvers.

\begin{figure}[!htb]
\centering     
   \includegraphics[scale=0.4]{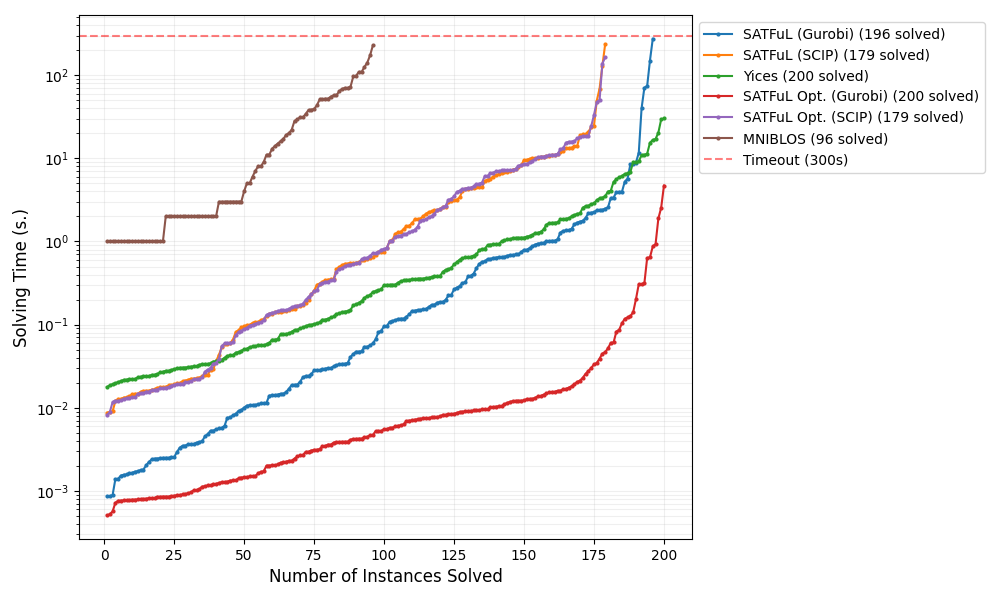}
   \caption{Cactus plot for product logic.}
   \label{fig:cactus plot product logic}
\end{figure}

\paragraph{Product Logic}
Product logic is the less explored fuzzy logic with respect to SAT solving. One reason for this is that its logical operators cannot be translated to Mixed Linear Programming problems. For instance, \textsf{FuzzyDL} does not provide support for product logic. 
In \cite{Vidal2016}, \textsf{MNiBLoS}, a SAT solver supporting several fuzzy logics and particularly product logic, was presented. \textsf{MNIBLOS} 
uses \textsf{Z3} theories. Furthermore, \textsf{Yices} supports non-linear arithmetic by means of the \textsf{MCSAT} engine \cite{}. We compared optimized and non-optimized \textsf{SATFuL} using \textsf{Gurobi} and \textsf{SCIP} with \textsf{MNiBLoS} and \textsf{Yices}.
Figure~\ref{fig:cactus plot product logic} shows a cactus plot depicting the results obtained
As shown in this figure,  optimized \textsf{SATFuL} using \textsf{Gurobi} offers the best performance among all solvers. Non-optimized \textsf{SATFuL} using \textsf{Gurobi} also performs well, but the optimized algorithm noticeably impacts the tool's performance. On the other hand, when using \textsf{SCIP} \textsf{SATFuL}  was able to solve most of the cases, but the optimization does not have a major impact in the performance of the algorithm. \textsf{MNiBLoS} has the worst performance among all solvers; it solved only 96 of 200 cases. 

Table~\ref{tab:solver_benchmark_product} shows the median running time for each solver, distinguishing each group of formulas. 

\begin{table}[htbp]
\centering
\caption{Median Running Times (seconds) Across Variable Counts by Solver}
\label{tab:solver_benchmark_columns}
\small
\setlength{\tabcolsep}{4pt} 
\begin{tabular}{c rrrrrr}
\toprule
\textbf{num\_vars} & \parbox{1.5cm}{\centering \textbf{SATFuL}\\ \textbf{Gurobi Opt}} & \parbox{1.4cm}{\centering \textbf{SATFuL}\\ \textbf{Gurobi}} & \textbf{Yices} & \parbox{1.4cm}{\centering \textbf{SATFuL}\\ \textbf{SCIP}} & \parbox{1.4cm}{\centering \textbf{SATFuL}\\ \textbf{SCIP Opt}} & \textbf{MNIBLOS} \\
\midrule
10 & 0.0009 & 0.0032 & 0.0295 &  0.0194 &  0.0193 &   2.0125 \\
20 & 0.0031 & 0.0706 & 0.1007 &  0.6032 &  0.5829 &  41.3008 \\
30 & 0.0083 & 0.4437 & 0.9275 &  2.7898 &  2.9706 & 300.0000 \\
40 & 0.0141 & 0.9943 & 1.6869 & 16.6961 & 17.7769 & 300.0000 \\
\bottomrule
\end{tabular}
\end{table}
As can be seen in the table, optimized \textsf{SATFuL} with Gurobi demonstrated decisive superiority, scaling well up to 40 variables ($0.0141\text{s}$) to achieve up to a $70\times$ speedup over standard \textsf{SATFuL} with \textsf{Gurobi} and a $119\times$ speedup over \textsf{Yices}. While  \textsf{Gurobi} backends consistently outperformed  \textsf{SCIP} configurations by an order of magnitude, the  optimizations provided negligible gains when paired with \textsf{SCIP}, highlighting a dependency on \textsf{Gurobi}'s internal solver architecture. Meanwhile, \textsf{Yices} maintained moderate competitive performance in middle tiers, whereas \textsf{MNiBLoS} suffered severe scalability bottlenecks, quickly hitting the 300-second timeout threshold at 30 and 40 variables.

Furthermore, a more detailed comparison between \textsf{SATFuL} and \textsf{Yices} (the two best solvers for this logic), shows that optimized \textsf{SATFuL} running \textsf{Gurobi} outperforms \textsf{Yices}. Particularly, for formulas with 40 variables.
\begin{figure}[!htb]
  \centering
  \makebox[\textwidth][c]{%
    \begin{subfigure}[b]{0.55\textwidth}
      \centering
      \includegraphics[width=\textwidth]{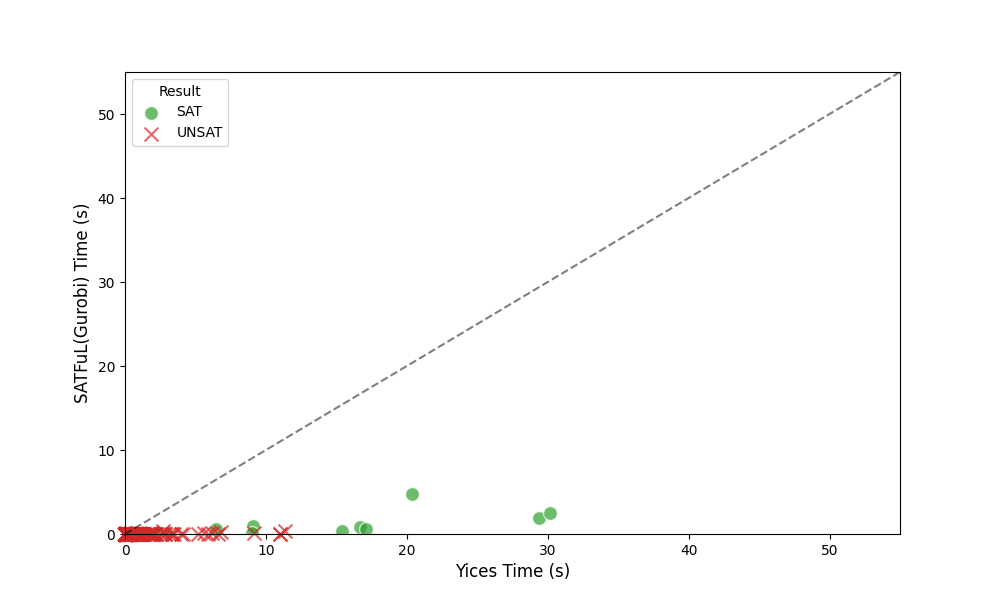}
      \caption{Cactus plot for product logic (SAT).}
      \label{fig:satful-vs-yices-prod-sat}
    \end{subfigure}%
    \hfill
    \begin{subfigure}[b]{0.55\textwidth}
      \centering
      \includegraphics[width=\textwidth]{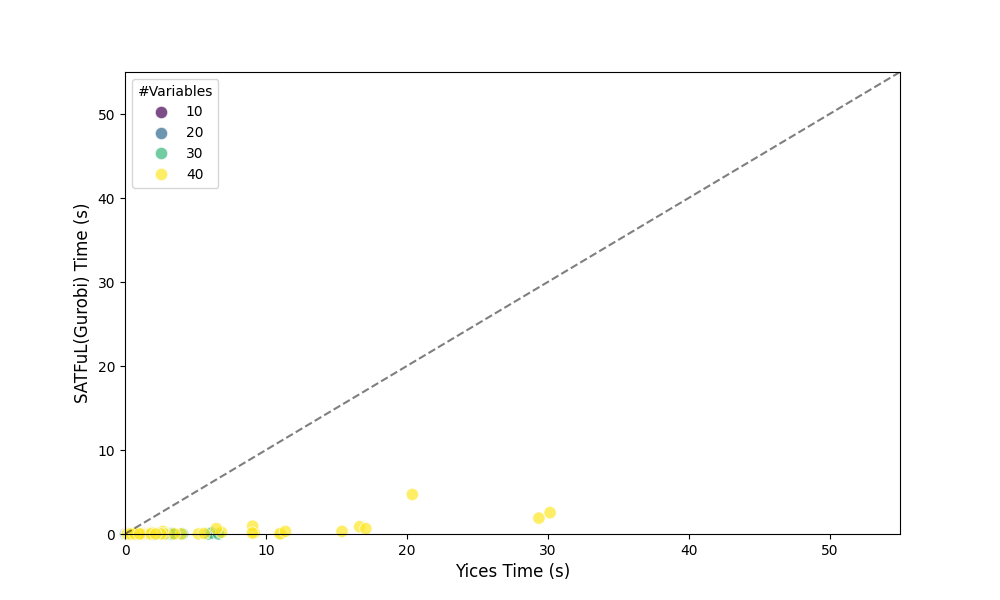}
      \caption{Cactus plot for product logic (Variables).}
      \label{fig:satful-vs-yices-prod-vars}
    \end{subfigure}%
  }
  \caption{Comparison of SATFuL (Gurobi) and Yices for product logic.}
  \label{fig:satful-vs-yices-prod}
\end{figure}

\section{Conclusions}
We introduced a SAT solving algorithm that uses mixed-integer non-linear programming targeting most relevant fuzzy logics; we proved the correctness of this algorithm, and proposed optimizations for product and G\"odel logics.
We have implemented a prototype of the fuzzy SAT solver: the tool \textsf{SATFuL}, an open-source Python-based SAT solver for fuzzy logics that translates fuzzy clauses into MINLP problems via a recursive algorithm. 
We have shown that \textsf{SATFuL} outperforms \textsf{MNiBLoS}, one of the few SAT solvers for  Product logic. Furthermore, for {\luk} logic, its performance is aligned with the state-of-the-art \textsf{fuzzySAT} solver  for SAT formulas, and outperforms this tool for UNSAT formulas. For G\"odel formulas, in which conditional states play a strong role, using the SMT solver \textsf{Yices} yields better performance,  even though \textsf{SATFuL} using \textsf{Gurobi} and the optimized  algorithm behaves better in some cases. Because of this, we have included the use of \textsf{Yices} as an option in \textsf{SATFuL}.

We aimed to provide a maintainable SAT solver for fuzzy logic that is easy to use and extend, following the successful path of Boolean SAT solvers. There are simple extensions for the SAT solver that were not included in this paper. For instance, adding support for fuzzy variables representing stochastic phenomena and relational operators. We leave them for future work.

%
%
%
\bibliography{biblio}

\end{document}